\documentclass[journal]{IEEEtran}

\usepackage{graphicx,color} %color for pdf_t
\usepackage{verbatim}
\usepackage[cmex10]{amsmath}  %cmex10 ensures that only type 1 fonts will be used
\usepackage{amssymb}

\newtheorem{theorem}{\textbf{Theorem}}

\newtheorem{remark}{Remark}

\begin{document}

%\runningheads{S.J. Johnson, L. Ong, C.M. Kellett}{Joint Channel-Network Coding}
%\categorytitle{Coding}

% \title{Joint Channel-Network Coding Strategies for Networks with Low Complexity Relays}
% \author{Sarah J. Johnson, Lawrence Ong\corrauth, and Christopher M. Kellett}
% \address{School of Electrical Engineering and Computer Science,
% University of Newcastle, Callaghan, NSW 2308, Australia.}
% \corraddr{School of Electrical Engineering and Computer Science,
% University of Newcastle, Callaghan, NSW 2308, Australia. E-mail: lawrence.ong@cantab.net\\
% This paper appeared in part at the Australasian Telecommunication Networks and Applications Conference
% (ATNAC), Christchurch, New Zealand, December 2007 and The IEEE
% Information Theory Workshop (ITW), Taormina 2009.\\
% This work is supported by the
% Australian Research Council under grants DP0877258 and DP0665742.
% }

%---------- Title ----------
\title{Joint Channel-Network Coding Strategies for Networks with Low Complexity Relays}
\author{Sarah J.\ Johnson, Lawrence Ong, and Christopher M.\ Kellett
\thanks{This paper appeared in part at the Australasian Telecommunication Networks and Applications Conference
(ATNAC), Christchurch, New Zealand, December 2007 and The IEEE
Information Theory Workshop (ITW), Taormina 2009.}
\thanks{This work is supported by the
Australian Research Council under grants DP0877258 and DP0665742.}
\thanks{The authors are with the School of Electrical Engineering and
Computer Science, University of Newcastle, Callaghan, NSW 2308,
Australia (email:\{sarah.johnson, chris.kellett\}@newcastle.edu.au, lawrence.ong@cantab.net).}
}

\maketitle

\begin{abstract}
  We investigate joint network and channel coding schemes for networks
  when relay nodes are not capable of performing
  channel coding operations.  Rather, channel encoding is performed at
  the source node while channel decoding is done only at the
  destination nodes.  We examine three different decoding strategies:
  independent network-then-channel decoding, serial network and channel decoding,
  and joint network and channel decoding.  Furthermore, we describe how to
  implement such joint network and channel decoding using iteratively
  decodable error correction codes.  Using simple networks as a model, we
  derive achievable rate regions and use simulations to demonstrate the effectiveness of the three decoders.
\end{abstract}

% \begin{keywords}
% Network coding, Channel coding, Capacity, LDPC Codes
% \end{keywords}

\maketitle

\section{Introduction}

Classically, communication over a network involves network nodes
whose sole function is the routing of packets.  Recently, however,
Ahlswelde et al.~\cite{ACLY00-TIT} observed that by allowing the
intermediate network nodes to combine information, a greater network
throughput can be obtained. This strategy is referred to as network
coding.

In much of the literature on network coding, each link is assumed to
have its own independent channel coding system and hence each link is
assumed to be error-free.  Indeed, for certain independent memoryless
networks, separating the channel and network coding in this way
guarantees asymptotically optimal error correction
\cite{SYC06-TIT,Borade_NetworkIF}.  However, for other networks,
examples have been given showing that, in general, network and channel
coding must be performed jointly to achieve the best performance
\cite{EMHRKKH03-DIMACS}.

Several authors have investigated this form of combining channel and
network coding for the wireless relay channel.  Use of iteratively
decodable codes such as turbo codes \cite{Haus-ETT09}, \cite{HaHa06-ICC} and
low-density parity-check (LDPC) codes \cite{HaScOi-Allerton05},
\cite{CBSA-JSAC07}, \cite{HuDu-TWC07}, \cite{RaYuTIT07},
\cite{BaLi-ISIT06} are also common.  A common feature in all of
these schemes is that the relay node decodes each packet prior to
performing network coding. In addition, \cite{HaHa06-ICC} and
\cite{TNB08} include automatic repeat requests (ARQ) as another
layer of error protection.

Recently, \cite{LSB-TIT09} and \cite{SKM} have investigated networks
where the network nodes have differing capabilities.  In particular,
\cite{SKM} considers a hierarchical network where sensors have limited
computing and communication capabilities and intermediate relay nodes,
which communicate to a central server or access point, are more
capable.  On the other hand, \cite{LSB-TIT09} looks to minimize the
capabilities required by network nodes, proposing networks where not
all nodes necessarily perform coding functions.

In this paper we investigate the combination of channel and network
coding in a simple cooperative network with noisy network links.
Similar in spirit to \cite{LSB-TIT09} and \cite{SKM}, we assume the
intermediate nodes have limited computing capabilities. In
particular, we assume the intermediate nodes do not perform channel
coding operations. Rather the nodes simply forward packets or
perform the operation of XOR'ing two incoming packets.  This
differs from the traditional network coding strategy of decoding
each packet at each node, XOR'ing the messages and then re-encoding
the result. In our networks we only perform end-to-end channel
coding, all channel encoding operations are performed at the source
and all decoding operations are done at the destination. We model
each link in the network as a binary symmetric channel as we assume
each node makes a hard decision on its received signals.

We investigate three decoding strategies: independent
network-then-channel decoding, serial network and channel decoding,
and joint network and channel decoding and illustrate these
strategies using LDPC codes. LDPC codes have been proposed for many
network based applications including relay-networks
\cite{HaScOi-Allerton05}, \cite{YangK07-RelayNC} and sensor networks
\cite{Yang_LDPCSensor07}. Note that, unlike the schemes in
\cite{HaScOi-Allerton05} and \cite{YangK07-RelayNC}, the messages in
our networks are only encoded by the channel code once at the
source(s) and decoded once at the destination node. Unlike the
schemes in \cite{Yang_LDPCSensor07} the sources are not correlated.

In Section~\ref{strategies:sec} we describe the three
above-mentioned decoding strategies. In
Section~\ref{Rate_regions:sec} we derive achievable rate regions for
each decoding strategy on a simple network to demonstrate the
advantage of joint decoding for cooperative networks with noisy
network links and end-to-end channel coding. In
Section~\ref{LDPC:sec} we describe how iterative decoders for
low-density parity-check (LDPC) can be constructed in practice for
each decoding strategy and provide simulation results showing the
benefits of the proposed joint network and channel decoding
strategy.

\section{Independent, Serial, and Joint Decoding} \label{strategies:sec}

Suppose the source(s) generate binary message vectors $\mathbf{u}_1,
\ldots, \mathbf{u}_S$ which are each encoded with the channel codes
$\mathcal{C}_1, \ldots, \mathcal{C}_S$, respectively (however the
same code can be used for some or all of the messages). For
simplicity, we will assume that all codes are of the same length and
each packet contains a single codeword. The generator matrices for
the codes $\mathcal{C}_1, \ldots, \mathcal{C}_S$ are given by $G_1,
\ldots,G_S$ respectively and so the codewords for messages
$\mathbf{u}_1, \ldots, \mathbf{u}_S$ are thus $\mathbf{c}_1 =
\mathbf{u}_1 G_1, \ldots, \mathbf{c}_S = \mathbf{u}_S G_S$,
respectively. The noisy versions of the codewords received at the
destination(s) are labeled $\widetilde{\mathbf{c}_1}, \ldots,
\widetilde{\mathbf{c}_S}$, respectively and we will write
$\widehat{\mathbf{c}_1}, \ldots, \widehat{\mathbf{c}_S}$ for the
decoded codewords. Packets which contain the modulo-2 sum of two or
more (noise-corrupted) codewords are produced by the low-complexity
intermediate network nodes which employ network coding to improve
the throughput of the network. We will write $\mathbf{{c}}_{i,j}$
for $\mathbf{{c}}_i \oplus \mathbf{{c}}_j$, where $\oplus$
represents a bit-wise XOR (or bit-wise modulo-2 addition), and thus
$\mathbf{\widetilde{c_{i,j}}}$ for the noisy received version of
$\mathbf{{c}}_{i,j}$. We assume the destination(s) know which
codebook $C_1$, ..., $C_S$ generated the original codewords and how
packets have been combined while traversing the network; e.g., via
the use of a packet header attached to each packet.

The aim of this paper is to decode the messages $\mathbf{u}_1,
\ldots, \mathbf{u}_S$ when the destination has noisy versions of one
or more of these combined packets and may also have noisy versions
of one or more packets containing original codewords. The approaches
we consider are independent network and channel coding, serial
network and channel coding and joint network and channel coding.

\subsection{Independent network-then-channel decoding}

The throughput benefit of the network code will be realized simply
by performing network decoding at the destination to recover noisy
versions of the transmitted codewords before independently decoding
each codeword with its corresponding error correction code.

For example, a destination node which receives
$\widetilde{\mathbf{c}_{1}}$, $\widetilde{\mathbf{c}_{2}}$ and
$\widetilde{\mathbf{c}_{2,3}}$ on three incoming links will
calculate $\widetilde{\mathbf{c}_{3}} = \widetilde{\mathbf{c}_2}
\oplus \widetilde{\mathbf{c}_{2,3}}$. The original messages can then
be found by using channel decoding on $\widetilde{\mathbf{c}_{1}}$
to obtain $\widehat{\mathbf{c}_{1}}$ and separately decoding
$\widetilde{\mathbf{c}_{2}}$ to obtain $\widehat{\mathbf{c}_{2}}$
and $\widetilde{\mathbf{c}_{3}}$ to obtain
$\widehat{\mathbf{c}_{3}}$. All three channel decoders can be run in
parallel to improve the speed of the decoding at the destination.
However, this strategy is clearly suboptimal as the noise in
$\widetilde{\mathbf{c}_2}$ will be carried into the calculation of
$\widetilde{\mathbf{c}_{3}}$.

\subsection{Serial network and channel decoding}

A potential improvement over independent network and channel coding
is to employ serial decoding, where decoding is performed on the
packet received on one of the incoming links and the decoded
information from this first link is shared with the decoders for the
packets received on the remaining links. This is repeated one link
at a time until the all of the packets are decoded.

The obvious strategy is that channel decoding is first performed on
the packets corresponding to an original codeword. Network decoding
is then applied using the decoded codewords and the received
combined packets (i.e., network-coded packets) to obtain noisy
versions of the remaining codewords. These are then decoded with
their respective channel decoder.

For example, a destination which receives
$\widetilde{\mathbf{c}_{1,2}}$, $\widetilde{\mathbf{c}_{2}}$ and
$\widetilde{\mathbf{c}_{1,3}}$ will decode
$\widetilde{\mathbf{c}_{2}}$ to obtain $\widehat{\mathbf{c}_2}$,
then calculate $\widetilde{\mathbf{c}_{1}} = \widehat{\mathbf{c}_2}
\oplus \widetilde{\mathbf{c}_{1,2}}$. Then
$\widetilde{\mathbf{c}_{1}}$ is decoded by its channel decoder to
obtain $\widehat{\mathbf{c}_1}$ and $\widetilde{\mathbf{c}_{3}} =
\widehat{\mathbf{c}_1} \oplus \widetilde{\mathbf{c}_{1,3}}$. Finally
$\widetilde{\mathbf{c}_{3}}$ is separately decoded by its channel
decoder to obtain $\widehat{\mathbf{c}_3}$. This method can improve
the performance of the channel decoding over independent schemes, as
we will see in Sections~\ref{Model:sec} and \ref{Rate_regions:sec},
but increases the decoding delay at the destination since the
channel decoders are run serially. In this example three decoder
applications are required in series but the extra delay will only
grow as the principle is extended to more complex networks.

\subsection{Joint network and channel decoding by defining a joint code}

In joint network and channel decoding we define a single error
correction code which incorporates the structure in the individual
channel codes and the structure in the network. In
Section~\ref{LDPC:sec} we will see two ways to achieve this using
LDPC codes, by defining a code on the joint codeword consisting of
all received packets, or the joint codeword consisting of all
transmitted and all received packets.

\section{Capacity and Achievable Rate Regions} \label{Rate_regions:sec}

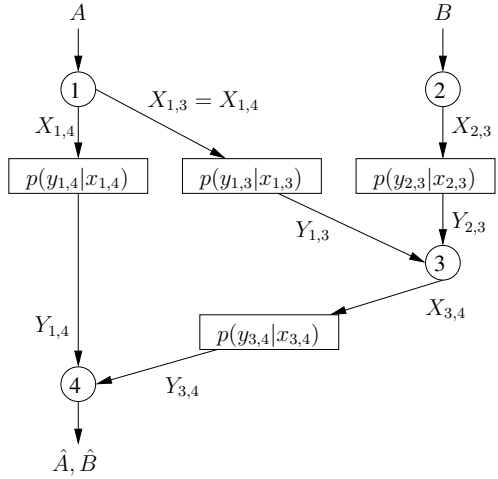
\begin{figure}
\centering
\resizebox{6.5cm}{!}{\begin{picture}(0,0)%
\includegraphics{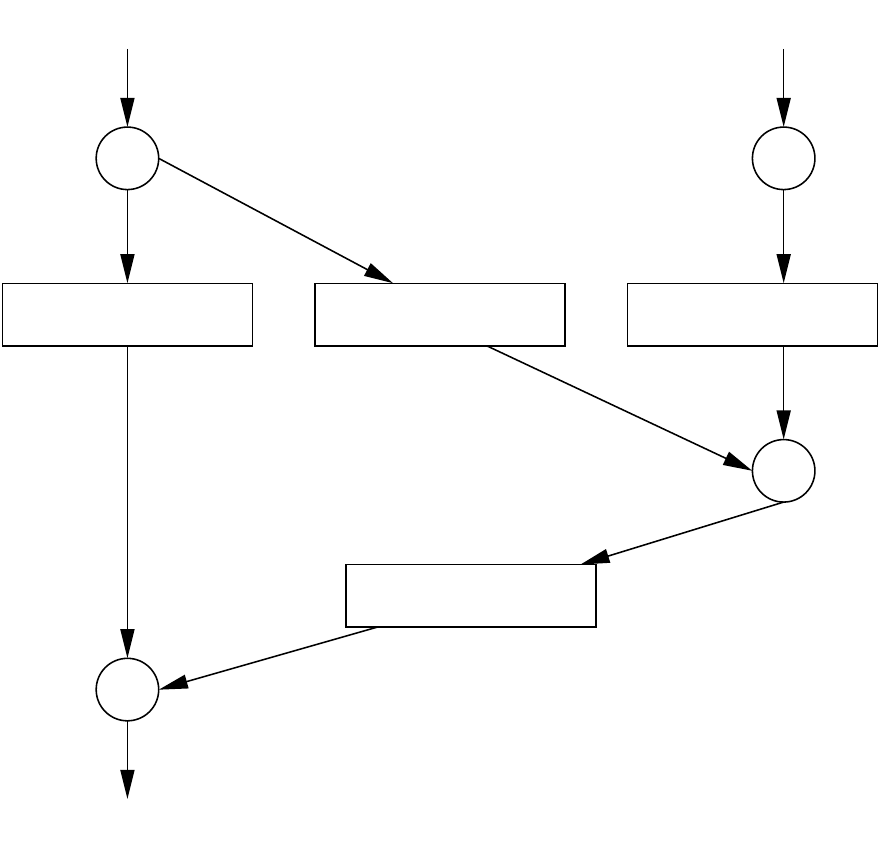}%
\end{picture}%
\setlength{\unitlength}{3947sp}%
\begingroup\makeatletter\ifx\SetFigFont\undefined%
\gdef\SetFigFont#1#2#3#4#5{%
  \fontsize{#1}{#2pt}%
  \fontfamily{#3}\fontseries{#4}\fontshape{#5}%
  \selectfont}%
\fi\endgroup%
\begin{picture}(4224,4120)(439,-3284)
\put(976,689){\makebox(0,0)[lb]{\smash{{\SetFigFont{12}{14.4}{\familydefault}{\mddefault}{\updefault}{\color[rgb]{0,0,0}$A$}%
}}}}
\put(4126,689){\makebox(0,0)[lb]{\smash{{\SetFigFont{12}{14.4}{\familydefault}{\mddefault}{\updefault}{\color[rgb]{0,0,0}$B$}%
}}}}
\put(826,-3211){\makebox(0,0)[lb]{\smash{{\SetFigFont{12}{14.4}{\familydefault}{\mddefault}{\updefault}{\color[rgb]{0,0,0}$\hat{A},\hat{B}$}%
}}}}
\put(976, 14){\makebox(0,0)[lb]{\smash{{\SetFigFont{12}{14.4}{\familydefault}{\mddefault}{\updefault}{\color[rgb]{0,0,0} 1}%
}}}}
\put(4126, 14){\makebox(0,0)[lb]{\smash{{\SetFigFont{12}{14.4}{\familydefault}{\mddefault}{\updefault}{\color[rgb]{0,0,0} 2}%
}}}}
\put(4126,-1486){\makebox(0,0)[lb]{\smash{{\SetFigFont{12}{14.4}{\familydefault}{\mddefault}{\updefault}{\color[rgb]{0,0,0} 3}%
}}}}
\put(976,-2536){\makebox(0,0)[lb]{\smash{{\SetFigFont{12}{14.4}{\familydefault}{\mddefault}{\updefault}{\color[rgb]{0,0,0} 4}%
}}}}
\put(676,-286){\makebox(0,0)[lb]{\smash{{\SetFigFont{12}{14.4}{\familydefault}{\mddefault}{\updefault}{\color[rgb]{0,0,0}$X_{1,4}$}%
}}}}
\put(1651,-61){\makebox(0,0)[lb]{\smash{{\SetFigFont{12}{14.4}{\familydefault}{\mddefault}{\updefault}{\color[rgb]{0,0,0}$X_{1,3}=X_{1,4}$}%
}}}}
\put(4276,-286){\makebox(0,0)[lb]{\smash{{\SetFigFont{12}{14.4}{\familydefault}{\mddefault}{\updefault}{\color[rgb]{0,0,0}$X_{2,3}$}%
}}}}
\put(4276,-1111){\makebox(0,0)[lb]{\smash{{\SetFigFont{12}{14.4}{\familydefault}{\mddefault}{\updefault}{\color[rgb]{0,0,0}$Y_{2,3}$}%
}}}}
\put(4051,-1861){\makebox(0,0)[lb]{\smash{{\SetFigFont{12}{14.4}{\familydefault}{\mddefault}{\updefault}{\color[rgb]{0,0,0}$X_{3,4}$}%
}}}}
\put(1801,-2536){\makebox(0,0)[lb]{\smash{{\SetFigFont{12}{14.4}{\familydefault}{\mddefault}{\updefault}{\color[rgb]{0,0,0}$Y_{3,4}$}%
}}}}
\put(2251,-2086){\makebox(0,0)[lb]{\smash{{\SetFigFont{12}{14.4}{\familydefault}{\mddefault}{\updefault}{\color[rgb]{0,0,0}$p(y_{3,4}|x_{3,4})$}%
}}}}
\put(601,-736){\makebox(0,0)[lb]{\smash{{\SetFigFont{12}{14.4}{\familydefault}{\mddefault}{\updefault}{\color[rgb]{0,0,0}$p(y_{1,4}|x_{1,4})$}%
}}}}
\put(2101,-736){\makebox(0,0)[lb]{\smash{{\SetFigFont{12}{14.4}{\familydefault}{\mddefault}{\updefault}{\color[rgb]{0,0,0}$p(y_{1,3}|x_{1,3})$}%
}}}}
\put(3601,-736){\makebox(0,0)[lb]{\smash{{\SetFigFont{12}{14.4}{\familydefault}{\mddefault}{\updefault}{\color[rgb]{0,0,0}$p(y_{2,3}|x_{2,3})$}%
}}}}
\put(2926,-1186){\makebox(0,0)[lb]{\smash{{\SetFigFont{12}{14.4}{\familydefault}{\mddefault}{\updefault}{\color[rgb]{0,0,0}$Y_{1,3}$}%
}}}}
\put(676,-2011){\makebox(0,0)[lb]{\smash{{\SetFigFont{12}{14.4}{\familydefault}{\mddefault}{\updefault}{\color[rgb]{0,0,0}$Y_{1,4}$}%
}}}}
\end{picture}%
}
\caption{Channel inputs, channel outputs, and channel transition
probabilities} \label{fig:input-output-probability}
\end{figure}

In this section we analyze the achievable rates (in the Shannon
sense) of the decoding strategies, without any assumption on the
code structure, meaning that the channel codes are not necessarily
LDPC. We consider the network depicted in
Fig.~\ref{fig:input-output-probability} as the simplest network
which combines both network and error correction coding. The
technique presented in this section can be easily extended to more
complex networks.

Each link $i \rightarrow j$ from node $i$ to node $j$, is a binary
symmetric channel (BSC) with crossover probability $p_{i,j}$. The
packet transmitted over link $i \rightarrow j$, is
$\mathbf{X}_{i,j}$, and the received vector is denoted
$\mathbf{Y}_{i,j}$. We assume that the channels for the links are
independent, time invariant, and memoryless. For the BSC $i
\rightarrow j$, the transition probability function is given by
\begin{subequations}
\begin{align}
\Pr\{ Y_{i,j} = m | X_{i,j} = m \} & = 1 - p_{i,j},  \quad\quad m \in \{0,1\}\\
\Pr\{ Y_{i,j} = n | X_{i,j} = m \} & = p_{i,j}, \quad m,n \in
\{0,1\}, n \neq m.
\end{align}
\end{subequations}
In words, with probability $p_{i,j}$ the input symbol is received in
error. Equivalently, we can write
\begin{equation}
Y_{i,j} = X_{i,j} \oplus E_{i,j}
\end{equation}
where $E_{i,j} \in \{0,1\}$ and $\Pr\{E_{i,j}=1\} = p_{i,j}$.

In this network, nodes 1 and 2 are sources for messages $A$ and $B$
respectively, but with the constraint that $X_{1,3}=X_{1,4}$. This
captures the fact that all source messages are encoded only once, at
their respective source nodes, and are not decoded (and re-encoded)
except at the destination. We let $A$ and $B$ be independently,
randomly, and uniformly chosen from the message alphabets
$\{1,2,\dotsc, \lfloor2^{nR_A}\rfloor\}$ and $\{1,2,\dotsc,
\lfloor2^{nR_B}\rfloor\}$ respectively, where $n$ is the block
length of the channel codes for all channels. The aim is to send
both $A$ and $B$ to node 4 in $n$ channel uses on each link. We use
$\hat{A}$ and $\hat{B}$ to denote the estimates for $A$ and $B$
respectively at node 4. The rate pair $(R_A,R_B)$ is
\emph{achievable} if $\Pr\{(\hat{A},\hat{B}) \neq (A,B)\}$ can be
made arbitrarily small. Node 4 can \emph{reliably} decode $A$ (or
$B$) iff $R_A$ (or $R_B$) is achievable. The \emph{capacity} is
defined as the set of all achievable rates.

It can be easily shown that the capacity of the BSC $i \rightarrow
j$ is
\begin{equation}
\mathbb{C}_{i,j} = 1 - H(p_{i,j}),
\end{equation}
where $H(p_{i,j}) = -p_{i,j} \log p_{i,j} -
(1-p_{i,j}) \log (1-p_{i,j})$. The capacity is achieved with an
equiprobable channel input distribution $p(x_{i,j})$.

\emph{Transmission at sources:} Nodes 1 and 2 send codewords for
messages $A$ and $B$ respectively: node 1 sends $\mathbf{X}_{1,3}(A)
\in \{0,1\}^n$ and $\mathbf{X}_{1,4}(A) \in \{0,1\}^n$ on links $1
\rightarrow 3$ and $1 \rightarrow 4$ respectively, where
$\mathbf{X}_{1,3}(A)=\mathbf{X}_{1,4}(A)$; and node 2 sends
$\mathbf{X}_{2,3}(B) \in \{0,1\}^n$ on link $2 \rightarrow 3$.

\emph{Linear codes:} We assume that linear codes are used. It has been shown by Elias~\cite{elias55} that
the capacity of the BSC is achievable by linear codes.

\subsection{Independent Network-then-Channel Decoding}
\emph{Strategy:} Node 4 decodes $\mathbf{X}_{1,4}(A)$ from
$\mathbf{Y}_{1,4}$. Independently on the other link, it subtracts
$\mathbf{Y}_{1,4}$ from $\mathbf{Y}_{3,4}$ and then decodes
$\mathbf{X}_{2,3}(B)$.

\begin{theorem} The achievable rate region for independent network-then-channel decoding,
$\Lambda_{\text{nc}}$, is the convex hull of all $(R_A,R_B)$
satisfying
\begin{subequations}
\begin{align}
R_A &\leq \mathbb{C}_{1,4} \label{eq:bsc-indep-network-then-channel-ra}\\
R_B &\leq \mathbb{C}''.\label{eq:bsc-indep-network-then-channel-rb}
\end{align}
\end{subequations}
Here, $\mathbb{C}_{1,4} = 1 - H(p_{1,4})$ is the capacity of link $2
\rightarrow 6$, and $\mathbb{C}'' = 1 - H(p'')$ is the capacity of a
BSC with cross-over probability $p''$ given in \eqref{eq-p''}.
\end{theorem}

\emph{Proof:} As message $A$ is decoded from $\mathbf{Y}_{1,4}$, we
see a point-to-point BSC $X_{1,4} \rightarrow Y_{1,4}$. So, we have
\eqref{eq:bsc-indep-network-then-channel-ra}.

By subtracting $\mathbf{Y}_{1,4}$ from $\mathbf{Y}_{3,4}$, we get
\begin{equation}
\mathbf{Y}_{3,4} \oplus \mathbf{Y}_{1,4} = \mathbf{X}_{2,3}(B) \oplus \mathbf{E}'',
\end{equation}
where $\mathbf{E}'' = \mathbf{E}_{1,3} \oplus \mathbf{E}_{2,3}
\oplus \mathbf{E}_{3,4} \oplus \mathbf{E}_{1,4}$. This can be viewed
as an equivalent BSC $X_{2,3} \rightarrow Y$, where $Y=Y_{3,4}
\oplus Y_{1,4}$, with cross-over probability $p''$, where
\begin{eqnarray}
p'' =& & \hspace*{-0.2in}  \Pr \{ E'' = 1\}\nonumber\\ = & & \hspace*{-0.2in}  p_{1,3}(1-p_{2,3})(1-p_{3,4})(1-p_{1,4})\nonumber\\
& &  \hspace*{-0.3in} + \ (1-p_{1,3})p_{2,3}(1-p_{3,4})(1-p_{1,4})\nonumber\\
& &  \hspace*{-0.3in} + \ (1-p_{1,3})(1-p_{2,3})p_{3,4}(1-p_{1,4})\nonumber\\
& &  \hspace*{-0.3in} + \ (1-p_{1,3})(1-p_{2,3})(1-p_{3,4})p_{1,4}\nonumber\\
& &  \hspace*{-0.3in} + \ (1-p_{1,3})p_{2,3}p_{3,4}p_{1,4} + p_{1,3}(1-p_{2,3})p_{3,4}p_{1,4} \nonumber\\
& &  \hspace*{-0.3in} + \ p_{1,3}p_{2,3}(1-p_{3,4})p_{1,4} +
p_{1,3}p_{2,3}p_{3,4}(1-p_{1,4}).\label{eq-p''}
\end{eqnarray}
So, node 4 can reliably decode message $B$ from $\mathbf{Y}_{3,4}
\oplus \mathbf{Y}_{1,4}$ up to the rate in
\eqref{eq:bsc-indep-network-then-channel-rb}.

Now, since the rate pair $(R_A,R_B) =
(\mathbb{C}_{1,4},\mathbb{C}'')$ is achievable, so are the rate
pairs $(\mathbb{C}_{1,4},0)$ (by switching node 2 off),
$(0,\mathbb{C}'')$ (by switching node 1 off), and $(0,0)$ (by
switching both nodes 1 and 2 off). By time sharing among any three
of these rate pairs, any rate pair in the convex hull of $(R_A,R_B)$
satisfying \eqref{eq:bsc-indep-network-then-channel-ra} and
\eqref{eq:bsc-indep-network-then-channel-rb} is achievable. $\hfill
\null \blacksquare \null$

\begin{remark}
Note that the order of network and channel decoding can also be reversed to give an independent channel-then-network decoding. This strategy was considered in our previous work \cite{ongjohnsonkellett09itw}. In channel-then-network decoding, node 4 first performs channel decoding independently on links $1 \rightarrow 4$ and $3 \rightarrow 4$ to obtain $\mathbf{X}_{1,4}(A)$ and $\mathbf{X}(A,B)$
respectively, where $\mathbf{X}(A,B)$ is a codeword that is a function of the messages $A$ and $B$, which is a result of the bit-wise XOR operation performed at node 3. Node 4 then performs network decoding to obtain message $B$ from $\mathbf{X}_{1,4}(A)$ and $\mathbf{X}(A,B)$. This strategy is not considered in this paper as the codebook $\{\mathbf{X}(A,B)\}$ defined for the combined messages $A$ and $B$ is, in general, not guaranteed to have the properties required for efficient decoding for the LDPC implementation in Section \ref{LDPC:sec}.
\end{remark}

\subsection{Serial Network and Channel Decoding}
\emph{Strategy:} Node 4 first decodes message $A$ from link $1
\rightarrow 4$. It then reconstructs $\mathbf{X}_{1,3}(A)$ and
subtracts it from $\mathbf{Y}_{3,4}$ before decoding message $B$.

\begin{theorem} The achievable rate region for serial decoding,
$\Lambda_{\text{serial}}$, is the convex hull of all $(R_A,R_B)$
satisfying
\begin{subequations}
\begin{align}
R_A &\leq \mathbb{C}_{1,4} \label{eq:bsc-serial-ra}\\
R_B &\leq \mathbb{C}', \label{eq:bsc-serial}
\end{align}
\end{subequations}
where $\mathbb{C}' = 1 - H(p')$ is the capacity of a
BSC with cross-over probability $p'$ given in \eqref{eq:p}.
\end{theorem}

\emph{Proof:} As message $A$ is first decoded from
$\mathbf{Y}_{1,4}$, we get \eqref{eq:bsc-serial-ra}.

By subtracting $\mathbf{X}_{1,3}(A)$ from $\mathbf{Y}_{3,4}$, we get
\begin{equation}
\mathbf{Y}_{3,4} \oplus \mathbf{X}_{1,3}(A) = \mathbf{X}_{2,3}(B) \oplus \mathbf{E}',\label{eq:serial-eq}
\end{equation}
where $\mathbf{E}'=\mathbf{E}_{1,3} \oplus \mathbf{E}_{2,3} \oplus \mathbf{E}_{3,4}$. By doing this, we get an equivalent BSC with cross-over probability
\begin{multline}\label{eq:p}
\Pr\{E'=1\}=p' = [(1-p_{1,3})(1-p_{2,3})+p_{1,3}p_{2,3}](p_{3,4})\\ +
[p_{1,3}(1-p_{2,3}) + (1-p_{1,3})p_{2,3}](1-p_{3,4}).
\end{multline}
So, node 4 can reliably decode message
$B$ if \eqref{eq:bsc-serial} is satisfied. $\hfill \null
\blacksquare \null$

\subsection{Joint Network and Channel Decoding}
\emph{Strategy:} Messages $A$ and $B$ are jointly decoded from the
received messages $\mathbf{Y}_{1,4}$ and $\mathbf{Y}_{3,4}$.

\begin{theorem}The achievable rate region for joint decoding,
$\Lambda_{\text{joint}}$, is the convex hull of all $(R_A,R_B)$
satisfying
\begin{subequations}
\begin{align}
R_A &\leq \mathbb{C}_{1,4} + \mathbb{C}' - \mathbb{C}'' \label{eq:joint-region-1}\\
R_B &\leq \mathbb{C}'\label{eq:joint-region-2}\\
R_A + R_B &\leq \mathbb{C}_{1,4} + \mathbb{C}',
\label{eq:joint-region-3}
\end{align}
\end{subequations}
where $\mathbb{C}'' = 1 - H(p'')$ is the capacity of a
BSC with cross-over probability $p''$ given in \eqref{eq-p''}, and 
$\mathbb{C}' = 1 - H(p')$ is the capacity of a
BSC with cross-over probability $p'$ given in \eqref{eq:p}.
\end{theorem}

\emph{Proof:} By doing joint decoding, we see a multiple-access
channel~\cite{liao72,ahlswede74} from $X_{1}$ and $X_{2,3}$ to
$Y_4$, where $X_1 = X_{1,3} = X_{1,4}$ and $Y_4 =
(Y_{1,4},Y_{3,4})$. Hence, we have the following capacity region:
\begin{subequations}
\begin{align}
R_A & \leq I(X_1;Y_4|X_{2,3}) \label{eq:bsc-joint-1}\\
R_B & \leq I(X_{2,3};Y_4|X_1) \label{eq:bsc-joint-2}\\
R_A + R_B & \leq I(X_1,X_{2,3};Y_4),\label{eq:bsc-joint-3}
\end{align}
\end{subequations}
maximized over all possible $p(x_1,x_{2,3})$. The capacity region
can be attained by independent and equiprobable $X_1$ and $X_{2,3}$.

Evaluating the RHS of \eqref{eq:bsc-joint-1} gives
\begin{subequations}
\begin{align}
&I(X_{1,3},X_{1,4};Y_{1,4},Y_{3,4}|X_{2,3})\nonumber \\
&= H(Y_{1,4},Y_{3,4}|X_{2,3}) - H(Y_{1,4},Y_{3,4}|X_{2,3},X_{1,3},X_{1,4})\\
&= (1 + H(p'')) - (H(p_{1,4}) + H(p')) \\
&= \mathbb{C}_{1,4} + \mathbb{C}' - \mathbb{C}''
\end{align}
\end{subequations}
where $p''$ is given in \eqref{eq-p''}, and 
$p'$ is given in \eqref{eq:p}.

Next, evaluating the RHS of \eqref{eq:bsc-joint-2} gives
\begin{subequations}
\begin{align}
&I(X_{2,3};Y_{3,4},Y_{1,4}|X_{1,3},X_{1,4})\nonumber\\
&= I(X_{2,3};Y_{3,4}|X_{1,3},X_{1,4}) \label{eq:joint-2-1} \\
&= I(X_{2,3};Y_{3,4}|X_{1,3}) \\
&= \mathbb{C}'. \label{eq:joint-2-2}
\end{align}
\end{subequations}
\eqref{eq:joint-2-1} is because given $(X_{1,3},X_{1,4})$, $X_{2,3}$
and $Y_{1,4}$ are independent, as $Y_{1,4} - X_{1,4} -
(X_{1,3},X_{2,3})$ forms a Markov chain. \eqref{eq:joint-2-2}
follows from \eqref{eq:serial-eq} and \eqref{eq:bsc-serial}.

Finally, evaluating the RHS of \eqref{eq:bsc-joint-3} gives
\begin{subequations}
\begin{align}
&I(X_{1,4},X_{1,3},X_{2,3};Y_{1,4},Y_{3,4})\nonumber\\
&= H(Y_{1,4},Y_{3,4}) - H(Y_{1,4},Y_{3,4}|X_{1,4},X_{1,3},X_{2,3})\\
&= 2 - (H(p_{1,4} + H(p'))\\
&= \mathbb{C}_{1,4} + \mathbb{C}'.
\end{align}
\end{subequations}
$\hfill \null \blacksquare \null$

\subsection{Comparison}
\begin{theorem}\label{lem:bsc}
The achievable rate regions satisfy
\begin{subequations}
\begin{align}
\Lambda_{\text{nc}} & \subseteq \Lambda_{\text{serial}} \subseteq \Lambda_{\text{joint}}\\
%\Lambda_{\text{cn,diff}} \subseteq \Lambda_{\text{cn,same}} &
%\subseteq \Lambda_{\text{serial}} \subseteq \Lambda_{\text{joint}}.
\end{align}
\end{subequations}
\end{theorem}

{\it Proof:} %of theorem~\ref{lem:bsc}]
We can show that
\begin{equation}
p'' - p' = (1-2p_{1,3})(1-2p_{2,3})(1-2p_{3,4})p_{1,3} \geq 0.
\end{equation}
The inequality above is because $0 \leq p_{1,3},p_{2,3},p_{3,4} \leq
\frac{1}{2}$. It can be shown by induction that $0 \leq p',p'' \leq
\frac{1}{2}$. This means $H(p'') \leq H(p')$ and $\mathbb{C}'' \leq
\mathbb{C}'$. So, the constraint
\eqref{eq:bsc-indep-network-then-channel-rb} is at least as strict
as the constraint \eqref{eq:bsc-serial}, and hence
$\Lambda_{\text{nc}} \subseteq \Lambda_{\text{serial}}$.

Lastly, the constraint \eqref{eq:bsc-serial-ra} is at least as
strict as the constraint \eqref{eq:bsc-joint-1} because $\mathbb{C}'
\geq \mathbb{C}''$. Summing
\eqref{eq:bsc-serial-ra} and \eqref{eq:bsc-serial} gives
\eqref{eq:bsc-joint-3}. So, $\Lambda_{\text{serial}} \subseteq
\Lambda_{\text{joint}}$.  $\hfill \null \blacksquare \null$

\begin{figure}[t]
\centering
\resizebox{7cm}{!}{\begin{picture}(0,0)%
\includegraphics{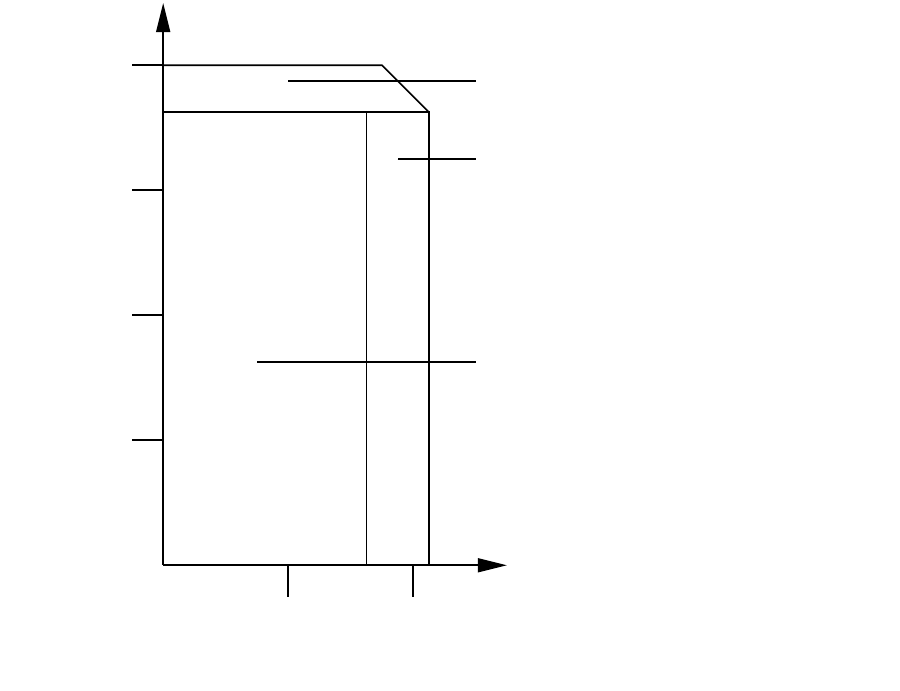}%
\end{picture}%
\setlength{\unitlength}{3947sp}%
\begingroup\makeatletter\ifx\SetFigFont\undefined%
\gdef\SetFigFont#1#2#3#4#5{%
  \reset@font\fontsize{#1}{#2pt}%
  \fontfamily{#3}\fontseries{#4}\fontshape{#5}%
  \selectfont}%
\fi\endgroup%
\begin{picture}(4311,3315)(718,-2164)
\put(1051,764){\makebox(0,0)[lb]{\smash{{\SetFigFont{12}{14.4}{\familydefault}{\mddefault}{\updefault}{\color[rgb]{0,0,0}0.8}%
}}}}
\put(1051,164){\makebox(0,0)[lb]{\smash{{\SetFigFont{12}{14.4}{\familydefault}{\mddefault}{\updefault}{\color[rgb]{0,0,0}0.6}%
}}}}
\put(1051,-436){\makebox(0,0)[lb]{\smash{{\SetFigFont{12}{14.4}{\familydefault}{\mddefault}{\updefault}{\color[rgb]{0,0,0}0.4}%
}}}}
\put(1051,-1036){\makebox(0,0)[lb]{\smash{{\SetFigFont{12}{14.4}{\familydefault}{\mddefault}{\updefault}{\color[rgb]{0,0,0}0.2}%
}}}}
\put(2551,-1861){\makebox(0,0)[lb]{\smash{{\SetFigFont{12}{14.4}{\familydefault}{\mddefault}{\updefault}{\color[rgb]{0,0,0}0.4}%
}}}}
\put(1951,-1861){\makebox(0,0)[lb]{\smash{{\SetFigFont{12}{14.4}{\familydefault}{\mddefault}{\updefault}{\color[rgb]{0,0,0}0.2}%
}}}}
\put(901,-1336){\rotatebox{90.0}{\makebox(0,0)[lb]{\smash{{\SetFigFont{12}{14.4}{\familydefault}{\mddefault}{\updefault}{\color[rgb]{0,0,0}$R_A$ [bits/channel use]}%
}}}}}
\put(1426,-2086){\makebox(0,0)[lb]{\smash{{\SetFigFont{12}{14.4}{\familydefault}{\mddefault}{\updefault}{\color[rgb]{0,0,0}$R_B$ [bits/channel use]}%
}}}}
\put(3076,689){\makebox(0,0)[lb]{\smash{{\SetFigFont{12}{14.4}{\familydefault}{\mddefault}{\updefault}{\color[rgb]{0,0,0}joint}%
}}}}
\put(3076,314){\makebox(0,0)[lb]{\smash{{\SetFigFont{12}{14.4}{\familydefault}{\mddefault}{\updefault}{\color[rgb]{0,0,0}serial}%
}}}}
\put(3076,-661){\makebox(0,0)[lb]{\smash{{\SetFigFont{12}{14.4}{\familydefault}{\mddefault}{\updefault}{\color[rgb]{0,0,0}network-then-channel}%
}}}}
\end{picture}%
}
\caption{Achievable rate
regions for different decoding schemes for BSC, $\rho_{i,j}=0.05$
for all links.}
\label{fig:compare-bsc-decoding-diff-code}
\end{figure}
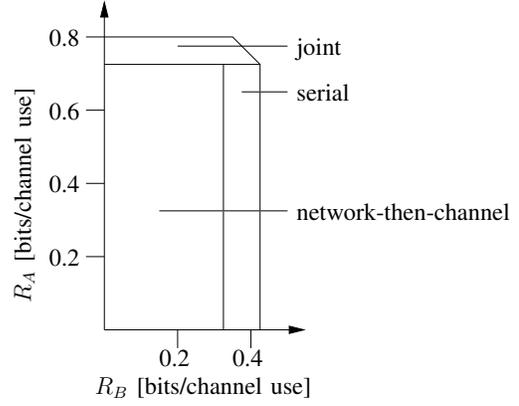

For example, Fig.~\ref{fig:compare-bsc-decoding-diff-code} %and
%\ref{fig:compare-bsc-decoding-same-code}
shows the achievable rates of the three decoding strategies when
$p_{i,j} = 0.05$ for all links.

In summary, we have the following comparison:
\begin{enumerate}
\item Serial decoding has the same rate region as network-then-channel
decoding for $R_A$ because both decodes $A$ from $\mathbf{Y}_{1,4}$. But serial decoding can improve $R_B$ over network-then-channel decoding as it subtracts a clean
(decoded) version of $\mathbf{X}_{1,3}(A)$ from the received message
$\mathbf{Y}_{3,4}$, and thus cancels the interference from message
$A$ before decoding message $B$. For network-then-channel decoding, a noisy version of $\mathbf{X}_{1,3}(A)$ is subtracted from $\mathbf{Y}_{3,4}$, and while the interference from message $A$ is removed, additional noise is also introduced at the same time.
% , and does so without restricting
%$R_A$ by requiring that $D$ be decoded directly.
\item Joint decoding can improve $R_A$ because node 4 decodes message $A$
from both $\mathbf{Y}_{1,4}$ and $\mathbf{Y}_{3,4}$ in joint
decoding but solely from $\mathbf{Y}_{1,4}$ in all other schemes.
Joint decoding does not improve $R_B$ over serial decoding as only
$\mathbf{Y}_{3,4}$ carries information about $B$. Upon canceling the
interference by $A$ when decoding $B$, serial decoding already
obtains the best rate region for $B$.
\end{enumerate}

The fact that serial decoding can improve $R_B$ over network-then-channel decoding, and that joint decoding can improve $R_A$ over both serial and network-then-channel decoding is also true for other channel models, for example the additive white Gaussian noise channel channel where each relay can only forward the summation of the signals it receives, scaled to account for constraint on the relay transmit power. %where each relay can only forward the scaled summation of the signals it receives.

While it is not surprising that joint decoding performs better than
serial decoding, which in turn outperforms independent decoding, the
rate region characterization allows us to analyze the improvement of
individual source data rates. It is interesting to see that
%independent channel-then-network decoding using the same codebook and
serial decoding is actually able to achieve a segment on the
capacity boundary in this example. This suggests that if a node in
the network only needs to decode data from selected sources, it may
not lose much performance, as far as achievable rate is concerned,
by considering the links independently. However, if the node is to
decode the data from all the sources, performing independent or
serial decoding results in a significant performance loss.

\section{Joint Network and Channel Coding Using Low-Density Parity-Check Codes} \label{LDPC:sec}

In this section we consider how to combine network and channel
coding using low-density parity-check codes and joint iterative
decoding. LDPC codes are block codes described by a sparse
parity-check matrix \cite{Gallager_LDPC_book,MacKay_sparse} first
presented by Gallager in 1962. Gallager proposed an iterative
decoding algorithm, now called sum-product decoding, which utilizes
the sparsity of the parity-check matrix to decode iteratively with
complexity linear in the code length. Using sum-product decoding,
LDPC codes have been shown to perform remarkably close to the
Shannon limit on many channels
\cite{Chung00045dB,MacKay_nearShannon}.

A length $n$ LDPC code is designed by specifying a sparse $m \times
n$ parity-check matrix $H$, and the code dimension is $k = n -
\mathrm{rank}_2(H)$. In most cases  $\mathrm{rank}_2(H) \approx m $
and $r = 1 - \frac{m}{n}$ is called the design rate. A generator
matrix for the code can be found using Gauss-Jordan elimination on
$H$ or encoding can be performed directly from $H$ in some cases.

An LDPC code is $(w_c,w_r)$-regular if all the columns of $H$ have
$w_c$ non-zero entries and all of the rows of $H$ have $w_r$
non-zero entries. A Tanner graph, \cite{Tanner}, displays the
relationship between codeword bits and parity checks in $H$. Each of
the $n$ code bits, and $m$ parity checks in $H$ are represented by a
vertex in the graph. A graph edge joins a code bit vertex to the
vertices of the parity checks that include it. A \emph{cycle} in a
Tanner graph is a sequence of connected code bits and parity checks
which start and end at the same vertex in the graph and contain no
other vertices more than once. The existence of cycles in the Tanner
graph are well known to hinder the performance of the sum-product
decoding algorithm (see e.g. \cite{MacKay_sparse}) and most LDPC
codes are designed to avoid cycles of size-4 (called 4-cycles) or
less.

We will propose two joint network and channel decoding strategies
which combine the parallel decoding advantages of independent
decoding and improve upon the error correction performance of serial
decoding by sharing error correction information between the channel
decoders.

In our first strategy we define a joint channel code which describes
the mapping of each transmitted message $\mathbf{u}_1, \ldots,
\mathbf{u}_S$ into each of the packets which have been received by
the destination. In effect we are incorporating the operations of
the network code into an extended channel code.

For example, a destination which receives
$\widetilde{\mathbf{c}_{1}}$ and $\widetilde{\mathbf{c}_{1,2}}$
(such as node 4 in the network depicted in
Fig.~\ref{fig:input-output-probability}) will define the generator
matrix for the code which maps $\mathbf{u}_{1}$, and
$\mathbf{u}_{2}$ to $ \mathbf{c}_{1}$ and $\mathbf{c}_{1} \oplus
\mathbf{c}_{2}$. For simplicity we will assume that the generator
and parity-check matrices are in standard form; i.e. the first $k$
columns of $G$ (respectively last $m=n-k$ columns of $H$) form a $k
\times k$ (respectively $m \times m$) identity matrix. However, the
resulting joint matrices apply equally for the non-systematic
parity-check matrices that are generally defined for LDPC codes.

Let
\[ G_1 = [I(k), \mathcal{G}_1] \;\;\;\; G_2 = [I(k), \mathcal{G}_2],
\]
\[ H_1 = [\mathcal{H}_1, I(m)] \;\;\;\; H_2 = [\mathcal{H}_2, I(m)],
\]
where $\mathcal{H}_1$ is the transpose of $\mathcal{G}_1$,
$\mathcal{H}_2$ is the transpose of $\mathcal{G}_2$ and $I(k)$ is
the $k \times k$ identity matrix, and both code rates are the same.
Consider a generator matrix $G_\mathrm{joint}$ for a code which
generates the codeword
\[\mathbf{c}_\mathrm{joint} = [\mathbf{c}_1,\mathbf{c}_1 \oplus
\mathbf{c}_2].\] The first $n$ bits in $\mathbf{c}_\mathrm{joint}$
are simply the codeword for $\mathbf{u}_1$ generated by
$\mathbf{u}_1 G_1$ and the second set of $n$ bits in
$\mathbf{c}_\mathrm{joint}$ are $\mathbf{u}_1 G_1 \oplus
\mathbf{u}_2 G_2$. Putting these equations in matrix form gives the
generator matrix:
\begin{eqnarray*} G_\mathrm{joint}  &=& \left[
  \begin{array}{cc}
     G_1 & G_1 \\
      0(k,n)   & G_2
 \end{array}
\right] \\ &=&  \left[
  \begin{array}{cccc}
     I(k) & \mathcal{G}_1 & I(k) & \mathcal{G}_1 \\
      0(k,k) & 0(k,m) & I(k) & \mathcal{G}_2
 \end{array}
\right],
\end{eqnarray*}
where $0(k,n)$ is the $k$ $\times$ $n$ all zeros matrix. We can then
write
\[ \mathbf{c}_\mathrm{joint} = [\mathbf{u}_1,  \mathbf{u}_2]G_\mathrm{joint}.
\]

$G_1$ and $G_2$ are already in standard form so to put
$G_\mathrm{joint}$ into standard form involves $k$ row operations
where the $j$-th row of $G_\mathrm{joint}$, for $1\leq j \leq k$, is
replaced by the modulo-2 sum of the $j$-th and $k+j$-th rows of
$G_\mathrm{joint}$ resulting in the matrix
\[ G'_\mathrm{joint} = \left[
  \begin{array}{cccc}
     I(k) & \mathcal{G}_1 & 0(k,k) & \mathcal{G}_1 \oplus \mathcal{G}_2 \\
      0(k,k) & 0(k,m) & I(k) & \mathcal{G}_2
 \end{array}
\right].
\] 

We can then define a joint network / channel parity-check matrix for
the network by
\begin{eqnarray*} H_\mathrm{joint} &=&
\left[
  \begin{array}{cccc}
     \mathcal{H}_1 & I(m) & 0(m,k) & 0(m,m)  \\
      \mathcal{H}_1 \oplus \mathcal{H}_2 & 0(m,m)  & \mathcal{H}_2 & I(m)
 \end{array}
\right] \\ &=& \left[
  \begin{array}{cc}
     H_1 & 0(m,n) \\
     H_1 \oplus H_2 & H_2
 \end{array}
\right]. \end{eqnarray*}

Then
\[ \mathbf{c}_\mathrm{joint} H_\mathrm{joint}^{T} = 0(1,2m),
\]
and we can jointly decode $\widetilde{\mathbf{c}_1}$ and
$\widetilde{\mathbf{c}_{1,2}}$ using $H_\mathrm{joint}$ to give
$\widehat{\mathbf{c}_1}$ and $\widehat{\mathbf{c}_{1,2}}$. The
decoded codeword for $\mathbf{u}_2$ is then simply \[
 \widehat{\mathbf{c}_{2}} = \widehat{\mathbf{c}_{1}}
\oplus \widehat{\mathbf{c}_{1,2}}.
\]
Note that the generator matrix $G_\mathrm{joint}$ was defined only
to motivate $H_\mathrm{joint}$, it will not be employed by the
source node which encodes $\mathbf{u}_1$ and $\mathbf{u}_2$
traditionally using $G_1$ and $G_2$. Importantly, $H_\mathrm{joint}$
is sparse when $H_1$ and $H_2$ are sparse so $H_\mathrm{joint}$
describes an LDPC code. Unlike independent and serial decoding,
joint decoding enables the decoder to use the information in
$\widetilde{\mathbf{c}_{1,2}}$ to decode $\mathbf{c}_{1}$.

If $H_1$ and $H_2$ are independent sparse parity-check matrices the
matrix $H_1 \oplus H_2$ will have many entries in common with $H_1$
(and $H_2$). This will lead to a significant number of 4-cycles in
the columns of $H_\mathrm{joint}$ which contain both $H_1 \oplus
H_2$ and $H_1$ (and the rows of $H_\mathrm{joint}$ which contain
both $H_1 \oplus H_2$ and $H_2$) and $4$-cycles are well known to
hinder the performance of the sum-product decoding algorithm (see
e.g. \cite{MacKay_sparse}).

To avoid these $4$-cycles we can design $H_1$ and $H_2$ so that the
$i$th column of $H_2$ contains all but one of its entries in common
with the $i$th column of $H_1$. However, this strategy is only
practical for networks with a limited number of channel codes. An
alternative strategy for joint decoding that avoids 4-cycles in the
joint Tanner graph is defined below.

For the special case where both messages are encoded with the same
code (i.e. $H_2 = H_1$) the joint parity-check matrix is
\[ H_\mathrm{joint} = \left[
  \begin{array}{cc}
     H_1 & 0(m,n) \\
     0(m,n) & H_1
 \end{array}
\right].
\]
The structure of $H_\mathrm{joint}$ reflects the fact that, since
$\mathbf{c}_1$ and $\mathbf{c}_2$ are both codewords of the linear
code represented by $H_1$, so too is $\mathbf{c}_1 \oplus
\mathbf{c}_2$. Thus, when the codes used for the messages are the
same, decoding with $H_\mathrm{joint}$ is actually independent
channel-then-network decoding rather than joint decoding.

\subsection{Joint network and channel decoding on an extended Tanner graph}

In this strategy a joint Tanner graph is defined for the channel and
error correction codes. The network coding operations are simply
modulo-2 sums of codewords and so can be considered as parity-check
equations which constrain the bits in the combined packets. The
extended Tanner graph includes the graphical representation of each
of the parity-check matrices $H_1, \ldots, H_S$ as well as bit nodes
for each of the combined packets, and constraint nodes for each of
the network coding operations.

For example, a destination which receives
$\widetilde{\mathbf{c}_{1}}$, $\widetilde{\mathbf{c}_{1,2}}$ and
$\widetilde{\mathbf{c}_{1,3}}$ will form a Tanner graph which
describes each of the parity check matrices $H_1$, $H_2$ and $H_3$,
includes bit nodes for all of the bits in $\mathbf{c}_{1,2}$ and
$\mathbf{c}_{1,3}$ and parity-check nodes for each of the network
coding operations
\[ \mathbf{{c}}_{1,2} = \mathbf{{c}}_1 \oplus \mathbf{{c}}_2, \]
\[ \mathbf{{c}}_{1,3} = \mathbf{{c}}_1 \oplus \mathbf{{c}}_3. \]

\begin{figure*}
  \begin{center}
    \includegraphics[width=10cm]{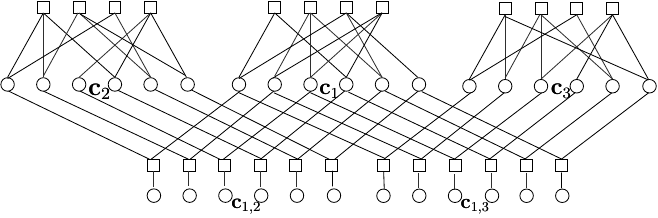}
    \caption{An extended Tanner graph. Three length six codewords $\mathbf{c}_{1}$, $\mathbf{c}_{2}$, and
$\mathbf{c}_{3}$ are generated by the source, and two combined
packets $\mathbf{{c}}_{1,2} = \mathbf{{c}}_1 \oplus \mathbf{{c}}_2$
and $\mathbf{{c}}_{1,3} = \mathbf{{c}}_1 \oplus \mathbf{{c}}_3$ are
generated by the network. \label{fig:Tanner}}
  \end{center}
\end{figure*}

Fig.~\ref{fig:Tanner} shows an extended Tanner graph at the
destination node which can be used to find $\mathbf{c}_{1}$,
$\mathbf{c}_{2}$, and $\mathbf{c}_{3}$ when
$\widetilde{\mathbf{c}_{1}}$, $\widetilde{\mathbf{c}_{1,2}}$ and
$\widetilde{\mathbf{c}_{1,3}}$ are received. The \emph{a priori} bit
LLRs for the bits not received directly by the destination node are
set to zero.

Different schedules can be used to decode the extended Tanner graph
but we will use a schedule of message passing decoding where one
iteration of the decoder corresponds to all of the bit nodes (for
the codewords and combined packets) updated in parallel and all of
the check nodes (for the channel codes and network codes) updated in
parallel. Note that this method of joint decoding for the butterfly
network was first presented in an earlier conference version of this
paper \cite{JoKe-ATNAC07} and independently, with slightly different
scheduling, in \cite{KZDL-ISIT08}.

\section{An Example - The Butterfly Network} \label{Model:sec}

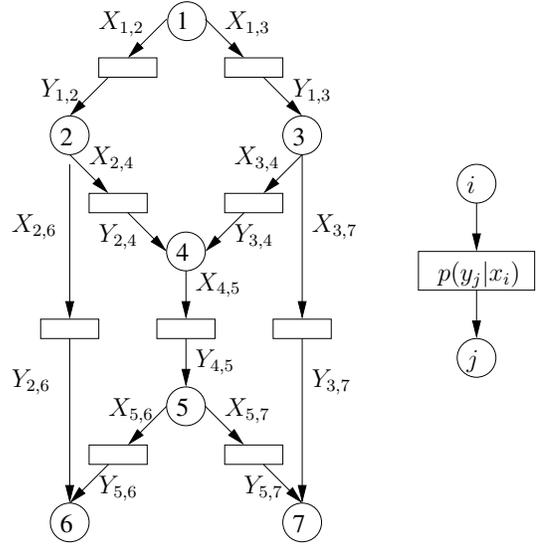
\begin{figure}
  \begin{center}
    \resizebox{7cm}{!}{\begin{picture}(0,0)%
\includegraphics{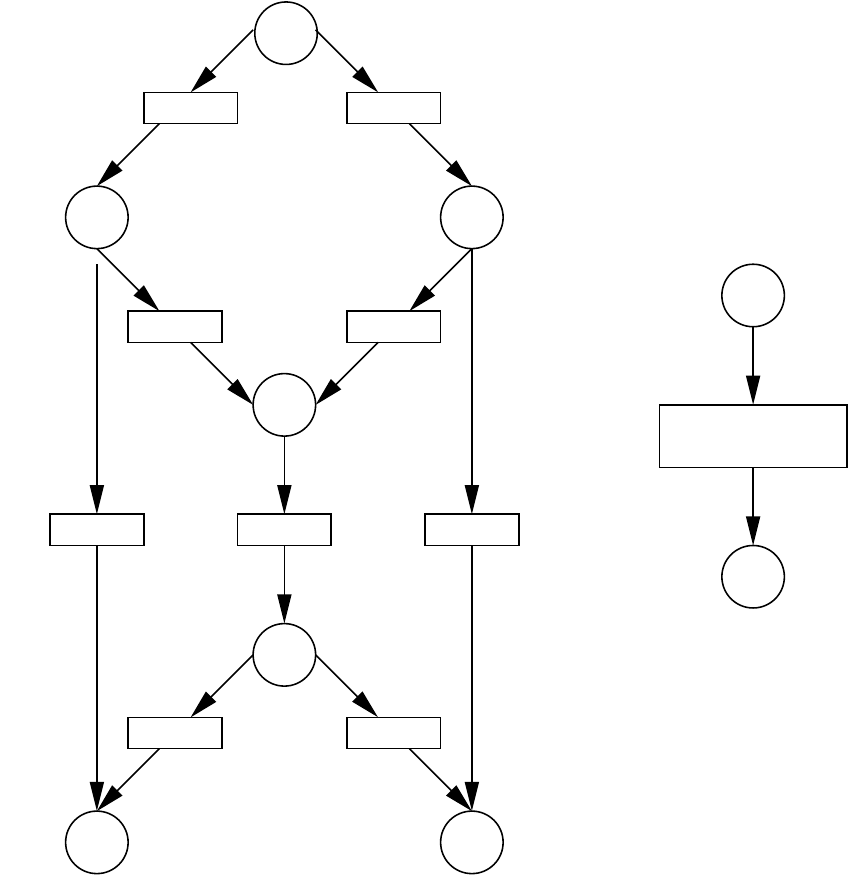}%
\end{picture}%
\setlength{\unitlength}{3947sp}%
\begingroup\makeatletter\ifx\SetFigFont\undefined%
\gdef\SetFigFont#1#2#3#4#5{%
  \reset@font\fontsize{#1}{#2pt}%
  \fontfamily{#3}\fontseries{#4}\fontshape{#5}%
  \selectfont}%
\fi\endgroup%
\begin{picture}(4077,4198)(1036,-3368)
\put(2326,-1186){\makebox(0,0)[lb]{\smash{{\SetFigFont{12}{14.4}{\familydefault}{\mddefault}{\updefault}{\color[rgb]{0,0,0} 4}%
}}}}
\put(2326,614){\makebox(0,0)[lb]{\smash{{\SetFigFont{12}{14.4}{\familydefault}{\mddefault}{\updefault}{\color[rgb]{0,0,0} 1}%
}}}}
\put(1426,-286){\makebox(0,0)[lb]{\smash{{\SetFigFont{12}{14.4}{\familydefault}{\mddefault}{\updefault}{\color[rgb]{0,0,0} 2}%
}}}}
\put(3226,-286){\makebox(0,0)[lb]{\smash{{\SetFigFont{12}{14.4}{\familydefault}{\mddefault}{\updefault}{\color[rgb]{0,0,0} 3}%
}}}}
\put(2326,-2386){\makebox(0,0)[lb]{\smash{{\SetFigFont{12}{14.4}{\familydefault}{\mddefault}{\updefault}{\color[rgb]{0,0,0} 5}%
}}}}
\put(1426,-3286){\makebox(0,0)[lb]{\smash{{\SetFigFont{12}{14.4}{\familydefault}{\mddefault}{\updefault}{\color[rgb]{0,0,0} 6}%
}}}}
\put(3226,-3286){\makebox(0,0)[lb]{\smash{{\SetFigFont{12}{14.4}{\familydefault}{\mddefault}{\updefault}{\color[rgb]{0,0,0} 7}%
}}}}
\put(1726,614){\makebox(0,0)[lb]{\smash{{\SetFigFont{12}{14.4}{\familydefault}{\mddefault}{\updefault}{\color[rgb]{0,0,0}$X_{1,2}$}%
}}}}
\put(1051,-961){\makebox(0,0)[lb]{\smash{{\SetFigFont{12}{14.4}{\familydefault}{\mddefault}{\updefault}{\color[rgb]{0,0,0}$X_{2,6}$}%
}}}}
\put(2701,614){\makebox(0,0)[lb]{\smash{{\SetFigFont{12}{14.4}{\familydefault}{\mddefault}{\updefault}{\color[rgb]{0,0,0}$X_{1,3}$}%
}}}}
\put(2776,-436){\makebox(0,0)[lb]{\smash{{\SetFigFont{12}{14.4}{\familydefault}{\mddefault}{\updefault}{\color[rgb]{0,0,0}$X_{3,4}$}%
}}}}
\put(1651,-436){\makebox(0,0)[lb]{\smash{{\SetFigFont{12}{14.4}{\familydefault}{\mddefault}{\updefault}{\color[rgb]{0,0,0}$X_{2,4}$}%
}}}}
\put(2476,-1411){\makebox(0,0)[lb]{\smash{{\SetFigFont{12}{14.4}{\familydefault}{\mddefault}{\updefault}{\color[rgb]{0,0,0}$X_{4,5}$}%
}}}}
\put(1276, 89){\makebox(0,0)[lb]{\smash{{\SetFigFont{12}{14.4}{\familydefault}{\mddefault}{\updefault}{\color[rgb]{0,0,0}$Y_{1,2}$}%
}}}}
\put(3226, 89){\makebox(0,0)[lb]{\smash{{\SetFigFont{12}{14.4}{\familydefault}{\mddefault}{\updefault}{\color[rgb]{0,0,0}$Y_{1,3}$}%
}}}}
\put(2776,-1036){\makebox(0,0)[lb]{\smash{{\SetFigFont{12}{14.4}{\familydefault}{\mddefault}{\updefault}{\color[rgb]{0,0,0}$Y_{3,4}$}%
}}}}
\put(1726,-1036){\makebox(0,0)[lb]{\smash{{\SetFigFont{12}{14.4}{\familydefault}{\mddefault}{\updefault}{\color[rgb]{0,0,0}$Y_{2,4}$}%
}}}}
\put(1051,-2161){\makebox(0,0)[lb]{\smash{{\SetFigFont{12}{14.4}{\familydefault}{\mddefault}{\updefault}{\color[rgb]{0,0,0}$Y_{2,6}$}%
}}}}
\put(3376,-961){\makebox(0,0)[lb]{\smash{{\SetFigFont{12}{14.4}{\familydefault}{\mddefault}{\updefault}{\color[rgb]{0,0,0}$X_{3,7}$}%
}}}}
\put(3376,-2161){\makebox(0,0)[lb]{\smash{{\SetFigFont{12}{14.4}{\familydefault}{\mddefault}{\updefault}{\color[rgb]{0,0,0}$Y_{3,7}$}%
}}}}
\put(4576,-661){\makebox(0,0)[lb]{\smash{{\SetFigFont{12}{14.4}{\familydefault}{\mddefault}{\updefault}{\color[rgb]{0,0,0} $i$}%
}}}}
\put(4576,-2011){\makebox(0,0)[lb]{\smash{{\SetFigFont{12}{14.4}{\familydefault}{\mddefault}{\updefault}{\color[rgb]{0,0,0} $j$}%
}}}}
\put(4351,-1336){\makebox(0,0)[lb]{\smash{{\SetFigFont{12}{14.4}{\familydefault}{\mddefault}{\updefault}{\color[rgb]{0,0,0}$p(y_j|x_i)$}%
}}}}
\put(1726,-2986){\makebox(0,0)[lb]{\smash{{\SetFigFont{12}{14.4}{\familydefault}{\mddefault}{\updefault}{\color[rgb]{0,0,0}$Y_{5,6}$}%
}}}}
\put(1801,-2386){\makebox(0,0)[lb]{\smash{{\SetFigFont{12}{14.4}{\familydefault}{\mddefault}{\updefault}{\color[rgb]{0,0,0}$X_{5,6}$}%
}}}}
\put(2851,-2986){\makebox(0,0)[lb]{\smash{{\SetFigFont{12}{14.4}{\familydefault}{\mddefault}{\updefault}{\color[rgb]{0,0,0}$Y_{5,7}$}%
}}}}
\put(2701,-2386){\makebox(0,0)[lb]{\smash{{\SetFigFont{12}{14.4}{\familydefault}{\mddefault}{\updefault}{\color[rgb]{0,0,0}$X_{5,7}$}%
}}}}
\put(2476,-2011){\makebox(0,0)[lb]{\smash{{\SetFigFont{12}{14.4}{\familydefault}{\mddefault}{\updefault}{\color[rgb]{0,0,0}$Y_{4,5}$}%
}}}}
\end{picture}%
}
    \caption{The butterfly network} \label{fig:butterfly-network}
  \end{center}
\end{figure}

In this example we will consider the butterfly network of
Fig.~\ref{fig:butterfly-network} (see, e.g.~\cite{ACLY00-TIT}). Each
link $i \rightarrow j$ from node $i$ to node $j$, is a binary
symmetric channel with crossover probability $p_{i,j}$. The source,
node $1$, generates two binary messages $\mathbf{u}_A$ and
$\mathbf{u}_B$. The codewords for messages $\mathbf{u}_A$ and
$\mathbf{u}_B$ are $\mathbf{c}_A = \mathbf{u}_A G_A$ and
$\mathbf{c}_B = \mathbf{u}_B G_B$, respectively. We also define a
code $\mathcal{C}_{AB}$ which consists of the set of codewords
\[\mathbf{c}_{AB} = \mathbf{c}_{A} \oplus \mathbf{c}_{B} \;\;\;\;
\forall \; \mathbf{c}_{A}, \mathbf{c}_{B}. \]

The codeword $\mathbf{c}_A$ is transmitted over link $1 \rightarrow
2$, i.e. $\mathbf{X}_{1,2} = \mathbf{c}_A$ and the codeword
$\mathbf{c}_B$ is transmitted over link $1 \rightarrow 3$, i.e.
$\mathbf{X}_{1,3} = \mathbf{c}_B$. Each of the nodes 2, 3, and 5
simply forward on the vector they receive, no processing is done of
any kind; i.e. $\mathbf{X}_{2,6} = \mathbf{X}_{2,4} =
\mathbf{Y}_{1,2}$ etc. Node 4 performs network coding by combining
the packets at its input
\begin{eqnarray*}
\mathbf{X}_{4,5} &=& \mathbf{Y}_{2,4} \oplus \mathbf{Y}_{3,4} \\
&=& \mathbf{c}_A \oplus \mathbf{E}_{1,2} \oplus  \mathbf{E}_{2,4} \oplus \mathbf{c}_B
\oplus \mathbf{E}_{1,3} \oplus  \mathbf{E}_{3,4}.
\end{eqnarray*}
No channel decoding is performed so $\mathbf{X}_{4,5}$ can be
thought of as a noisy version of the codeword $\mathbf{c}_{AB}$.

The destination, node $6$, knows which channel codes have been used
and has available $\mathbf{Y}_{2,6} = \widetilde{\mathbf{c}_A}$, a
noisy version of $\mathbf{c}_A$, and $\mathbf{Y}_{5,6} =
\widetilde{\mathbf{c}_{AB}}$,  a noisy version of the codeword
$\mathbf{c}_{AB}$. (Although we only consider node $6$, an identical
argument applies to the node $7$.) As the nodes can only re-transmit
the binary vector detected at their input (or XOR two such binary
vectors) errors added by the links will occur as flipped bits. Thus
for networks which transmit over more general memoryless channels
the network can still be modeled using binary symmetric channels for
the links.

We can define a joint network / channel parity-check matrix for the
butterfly network between nodes $1$ and $6$ by
\begin{eqnarray*} H_\mathrm{joint} &=&
\left[
  \begin{array}{cc}
     H_A & 0(m,n) \\
     H_A \oplus H_B & H_B
 \end{array}
\right], \end{eqnarray*} and so we can jointly decode
$\widetilde{\mathbf{c}_A}$ and $\widetilde{\mathbf{c}_{AB}}$ using
$H_\mathrm{joint}$ to give $\widehat{\mathbf{c}_A}$ and
$\widehat{\mathbf{c}_{AB}}$. The decoded codeword is then simply
\begin{equation} \label{joint_B} \widehat{\mathbf{c}_{B}} = \widehat{\mathbf{c}_{A}}
\oplus \widehat{\mathbf{c}_{AB}}.
\end{equation}

%\subsubsection{Extended joint network and channel decoding}

For the extended joint network and channel coding we define the
extended codeword as the concatenation of the codewords
$\mathbf{c}_A$, $\mathbf{c}_B$, and $\mathbf{c}_{AB}$:
\[\mathbf{c}_\text{extn} = [\mathbf{c}_A,\mathbf{c}_B,\mathbf{c}_A \oplus
\mathbf{c}_B]. \]

It is easy to see that such codewords must satisfy the parity-check
matrix
\[ H_\text{extn} = \left[
  \begin{array}{ccc}
        H_A & 0(m,n) & 0(m,n)  \\
     0(m,n) & H_B    & 0(m,n) \\
       I(n) & I(n)   & I(n)
 \end{array}
\right]. \] The relationship $\mathbf{c}_{AB} = \mathbf{c}_A \oplus
\mathbf{c}_B$ is represented by the last $n$ parity-check
constraints in $H_\text{extn}$. $H_\text{extn}$ will be $4$-cycle
free whenever $H_{A}$ and $H_{B}$ are designed to be $4$-cycle free.
Of course, the decoder at node $6$ does not have values for \emph{a
priori} input probabilities for $\mathbf{c}_B$ with which to perform
decoding using $H_\text{extn}$. This can be easily remedied by
passing in \emph{a priori} probabilities $p(c_B(i)=1) = p(c_B(i)=0)
= 0.5$ for these received bits.

Then
\[ \mathbf{c}_\text{extn} H_\text{extn}^{T} = 0(1,2m+n),
\]
and so we can jointly decode $\widetilde{\mathbf{c}_A}$ and
$\widetilde{\mathbf{c}_{AB}}$ using $H_\text{extn}$ to give
$\widehat{\mathbf{c}_A}$ and $\widehat{\mathbf{c}_{B}}$; i.e. with
this scheme the decoded packets $\mathbf{c}_A$ and $\mathbf{c}_B$
are returned directly by the joint decoder.

\subsection{Simulation results} \label{results:sec}

Different randomly constructed ($3,6$)-regular rate-1/2 LDPC codes
free of 4-cycles (see e.g. \cite{MacKay_sparse}, and code from
\cite{Nealweb}) are used for the channel codes $\mathcal{C}_A$ and
$\mathcal{C}_B$ as they have a sparse parity-check matrix
representation with good sum-product decoding performances. Random
codes are chosen to focus on the decoding strategies rather than any
effects of a particular code design. We use codewords of length 500
bits and apply standard sum-product decoding with a maximum of 20
decoder iterations. For the independent and serial decoding schemes
this means a maximum of 20 iterations for each channel decoder, but
for the joint decoding schemes the single joint decoder uses a
maximum of 20 iterations.  So that each path is subject to roughly
the same level of noise, the links each have crossover probability
$p$ except for link $2 \rightarrow 6$ which has crossover
probability $3p$.

\begin{figure}[t!]
\begin{center}
\includegraphics[width = 8cm]{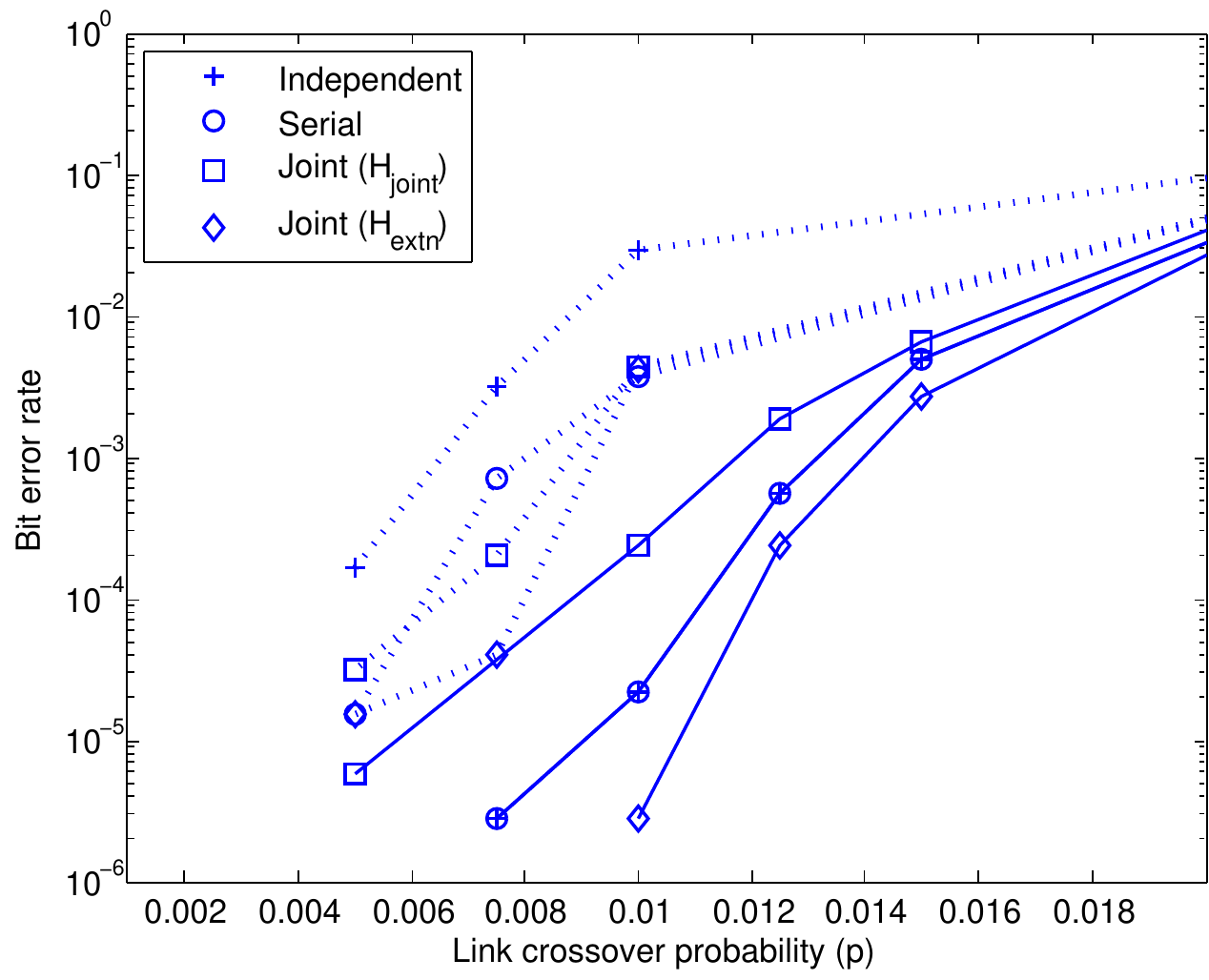}
\caption{Different length-500, rate-1/2, LDPC codes are used to
encode $\mathbf{u}_A$ and $\mathbf{u}_B$ which are transmitted over
the network in Fig.~\ref{fig:butterfly-network}. Shown is the bit
error rate performance of packet $A$ (solid curves) and packet $B$
(dotted curves) using the four different decoding schemes. The link
crossover probabilities are $p$ with the exception of link $2
\rightarrow 6$ which has crossover probability $3p$.
\label{fig:butterfly_diffH}}
\end{center}
\end{figure}

Fig.~\ref{fig:butterfly_diffH} shows the error correction
performance of the various decoding methods when $\mathbf{u}_A$ and
$\mathbf{u}_B$ are encoded with different randomly chosen LDPC
codes. We can see that independently decoding with the network and
then channel codes performs as expected, returning poor performances
for the decoding of $\mathbf{u}_B$, since it is corrupted by both
the errors on $\widetilde{\mathbf{c}_A}$ and those on
$\widetilde{\mathbf{c}_{AB}}$. Also as expected, using serial
decoding or either version of joint decoding,
$\widetilde{\mathbf{c}_{B}}$ is only corrupted by the errors from
$\widehat{\mathbf{c}_{A}}$ that remained after decoding and so the
decoding performance of $\mathbf{u}_B$ is significantly improved.

\begin{figure}[t!]
\begin{center}
\includegraphics[width = 8cm]{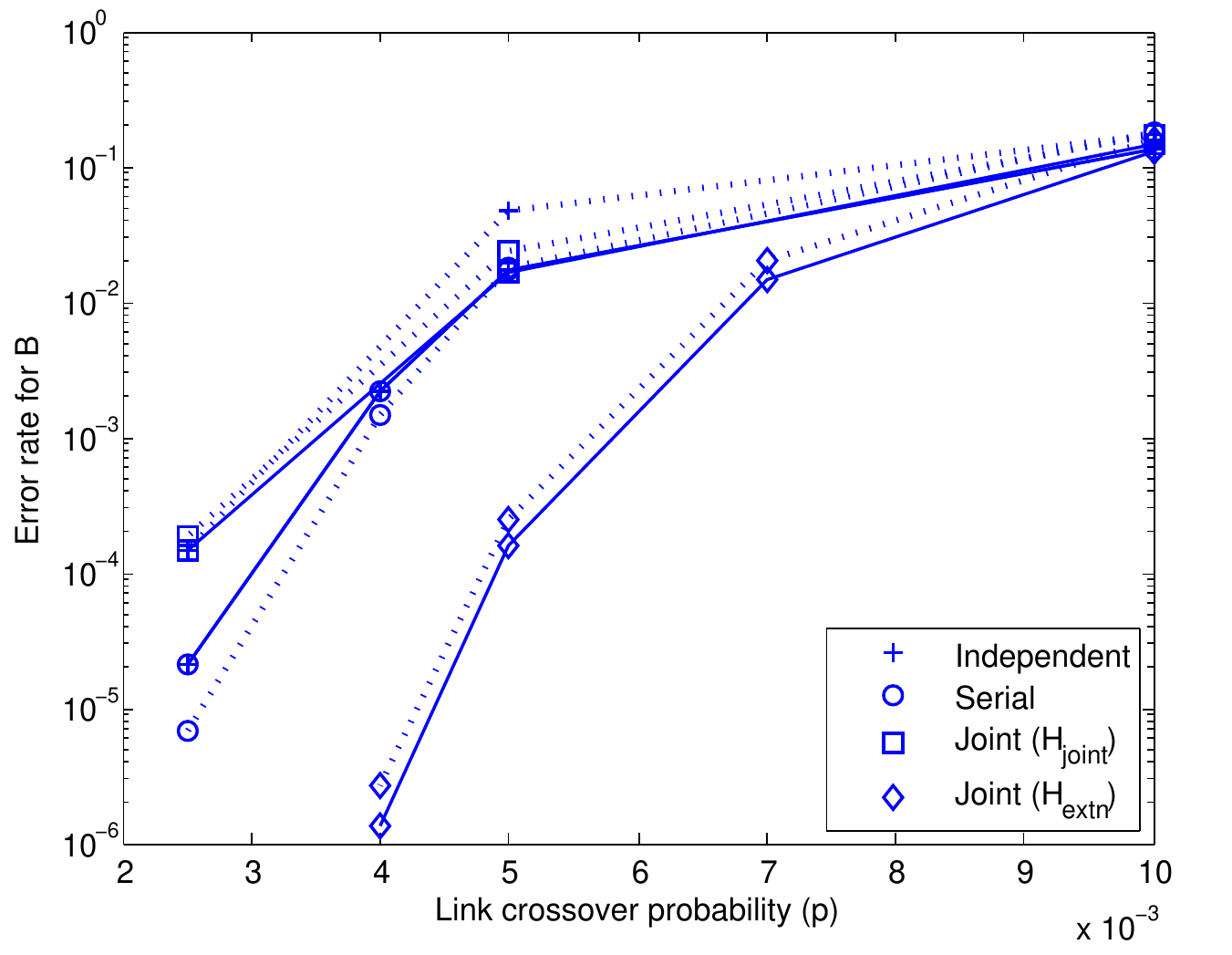}
\caption{Different length-500, rate-1/2, LDPC codes are used to
encode $\mathbf{u}_A$ and $\mathbf{u}_B$ which are transmitted over
the network in Fig.~\ref{fig:butterfly-network}. Shown is the bit
error rate performance of packet $A$ (solid curves) and packet $B$
(dotted curves) using the four different decoding schemes. The link
crossover probabilities are $p$ with the exception of link $2
\rightarrow 6$ which now has crossover probability $12p$.
\label{fig:butterfly_diffHpoor}}
\end{center}
\end{figure}

Fig.~\ref{fig:butterfly_diffHpoor} emphasizes the benefit that joint
decoding, by using the network code as part of a larger error
correction code, can provide to $\mathbf{u}_A$ over decoding with
the channel code alone. In this simulation the network is modified
to increase the crossover probability on link $2 \rightarrow 6$ to
be $12p$.

Overall, serial decoding performs equally as well as independent
network-then-channel decoding for $A$ but improves the performance
of $B$. The error rate of the first joint decoding scheme however,
is poorer than that of serial decoding for $A$ as it is hampered by
the 4-cycles in the parity-check matrix $H_{AB}$. The extended joint
decoding scheme, which is able to jointly decode $A$ and $B$ using a
cycle free Tanner graph, outperforms all the other schemes for both
$A$ and $B$.

The network coding performed at node 4 to improve the throughput of
the network has the unavoidable effect of reducing the BER
performance of packet $B$ (when compared to a network which uses two
channel transmissions at node 4 to send $A$ and $B$ separately to
node 5). However, by using joint decoding, this loss in performance
can be significantly reduced and, furthermore, the network coding
can even be used to improve the BER performance of packet $A$
(compared to the network without network coding).

Although joint decoding can not improve the rate region (in the
Shannon sense, i.e., rates with error probability approaching zero
using infinitely long code length) for $B$ over that of serial
decoding, simulation results show a bit error rate performance
improvement for $B$ when joint decoding with $H_\text{extn}$ is
used, showing that by jointly decoding $A$ and $B$ the convergence
performance of $B$ can be improved. Or put another way, if
$\mathbf{c}_{A}$ is received without error there will be no
improvement in the performance of joint over serial decoding of
$\mathbf{c}_{B}$. However, using a finite code length, where errors
occur in $\mathbf{c}_{A}$, joint decoding can remove more errors
from $\mathbf{c}_{A}$ so fewer errors remain to corrupt
$\mathbf{c}_{B}$. Furthermore, the decoding of $A$ and $B$ can be
successively improved iteratively using a better estimate of $A$ to
improve the decoding of $B$ and vice versa.

\subsection{Complexity}

The decoding schemes proposed here do not employ channel codes in
the intermediary network nodes, so the network complexity remains
the same for all decoding schemes. Sources $A$ and $B$ are encoded
independently by the encoder in the same way for each scheme and so
the source node complexity also does not change. The only increase
in complexity for the joint decoding schemes occurs for the decoder
at the destination node.

In general, the complexity of the sum-product decoding algorithm is
a linear function of the number of non-zero entries in the
parity-check matrix. For $(3,6)$-regular length $n$ LDPC codes the
total number of non-zero parity-check matrix entries in $H_A$ and
$H_B$ is $3n$ each. Thus, the independent and serial decoding
schemes have a total of $6n$ non-zero parity-check matrix entries
(over both matrices), while the joint network / channel decoding
matrix can have up to $6n$ additional non-zero entries (if there are
no entries overlapped in $H_A$ and $H_B$), and the extended joint
decoding matrix has $3n$ additional non-zero entries. Thus, using
rate half codes the number of non-zero parity-check matrix entries
is $6n$ for independent decoding, between $6n$ (common channel
codes) and $12n$ (completely disjoint parity-check matrices) for
joint decoding and $9n$ for extended joint decoding. For all the
schemes, decoding complexity remains linear in the block length, and
while the joint decoding schemes have a slightly higher decoding
complexity per iteration, their improved performance means that
fewer decoder iterations are actually required.

\section{Conclusion} \label{Conclusion:sec}

In this paper we have considered decoding schemes for low complexity
networks, where a message is encoded at the source node and decoded
at the destination node, and intermediate nodes perform network
coding operations, in our case modulo-2 addition, but no channel
coding. We have investigated three potential decoding schemes for
the destination: (1) independent decoding where the destination
decodes data from each link independently, (2) serial decoding where
the destination decodes data from each link independently, but in
series and by using the knowledge of previously decoded links, and
(3) joint decoding where the destination jointly decodes all data
from all the links simultaneously.

In networks with noisy links and low-complexity intermediary nodes
it can still be of benefit to perform a simple network coding
strategy, involving the XOR of (noise-corrupted) codewords rather
than messages, and use joint decoding at the destination to retrieve
the transmitted codewords. We saw that our proposed decoding scheme
improves the error correction performance of the network over the
independent and the serial decoding schemes, and does so without
adding network complexity. This is achieved by describing a joint
network code / channel code at the destination node and decoding
with iterative sum-product decoding. The new schemes require no
channel coding at the intermediary network nodes and only a small
amount of additional decoding complexity at the destination node.
The joint decoding produces improved error correction performances
and can significantly improve decoder speed.

\bibliographystyle{wileyj}

%\bibliography{../NetworkCoding.bib}

%\newpage

\end{document}